\documentclass[aps,prb,twocolumn,superscriptaddress,longbibliography]{revtex4-2}
\usepackage[colorlinks=true,linkcolor=blue,citecolor=blue]{hyperref}
\usepackage{amsfonts}
\usepackage{subfigure}
\usepackage{amsmath}
\usepackage{txfonts}
\usepackage{amssymb}
\usepackage{amsbsy}
\usepackage{epsfig}
\usepackage{graphicx}
\usepackage{epstopdf}
\usepackage{mathdots}
\usepackage{color}
\usepackage{cleveref}
\usepackage{bbm}
\usepackage{natbib}
\usepackage{url}
\usepackage[utf8]{inputenc}

\newcommand{\unit}{1\!\!1}

\newcommand{\be}{\begin{equation}}
\newcommand{\ee}{\end{equation}}

\newcommand{\bs}{\boldsymbol}

\begin{document}

%\title{Laser induced Flqouet topological phases in a 2D weak topological insulator with $\bs{\it unconventional}$ dispersion }
%\title{Role of a coupling term in a 2D Floquet topological insulator }
%\title{Construction-Dependent Floquet Topological Phases in Irradiated $N$-Stacked Su–Schrieffer–Heeger Chains}
\title{Construction-Dependent Floquet Topology in Circularly Driven $N$-Stacked Su-Schrieffer-Heeger Chains}

\author{Aayushi Agrawal}
\email{aagrawal@kias.re.kr}
\affiliation{School of Physics, Korea Institute for Advanced Study, Seoul 02455, South Korea}
\author{Jayendra N. Bandyopadhyay}
\email{jnbandyo@gmail.com}
\affiliation{Department of Physics, Birla Institute of Technology and Science, Pilani 333031, India}

\date{\today}

\begin{abstract}

We investigate construction-dependent Floquet topology in a two-dimensional lattice formed by stacking Su-Schrieffer-Heeger (SSH) chains with co-located sublattice orbitals. We present two constructions in the decoupled-chain limit. Construction-I: the constituent SSH chains are topologically trivial with winding number $w=0$. Construction-II: the constituent SSH chains are topologically nontrivial with $w=1$. Although these constructions are unitarily equivalent in the static limit, circular driving distinguishes them because the hopping processes related by the transformation correspond to different physical bond vectors and therefore acquire different Peierls phases. Using the Floquet-Sambe formalism, we determine the quasienergy spectra and phase diagrams in terms of the lower-band Chern number $C_L$ and directional Zak phases $(Z_x,\, Z_y)$. We find distinct Floquet phase structures, including low-frequency phases with $C_L = -3$ and $C_L = -4$ in the two constructions. Since the Chern number does not resolve the topology of the two quasienergy gaps separately, we further calculate the dynamical winding numbers $W_0$ and $W_\pi$, and verify the corresponding chiral edge states. In particular, at identical driving parameters, both constructions can have the same Chern numbers $C_L=-2$, while their gap-resolved winding numbers are $(W_0,\, W_\pi)=(0,\,+2)$ and $(-2,\,0)$, respectively. We also calculate the time-averaged optical Hall conductivity and show that $ C_Le^2/h$ determines its dc value for ideal full-band occupation. At the same time, its finite-frequency response contains photon-assisted contributions. Our results demonstrate that the physical construction of the constituent chains, although irrelevant to the static bulk spectrum, can play a decisive role in the Floquet topological structure. We discuss possible experimental realizations of the driven model.

\end{abstract}

\maketitle

\section{\label{Sec:Introduction}Introduction}

Periodic driving provides a route for modifying band structures and generating topological phases which are absent in their equilibrium counterparts \cite{Floquet-01,Floquet-top-01,Floquet-top-02, Floquet-top-03,Floquet-top-04,Floquet-top-05,Floquet-top-06, Basu-01, Basu-03}. For a Hamiltonian with period $T$, quasienergy is defined modulo the driving frequency $\Omega=2\pi/T$. A two-dimensional Floquet spectrum therefore contains not only a central gap at $\epsilon=0$, but also a distinct gap at the boundary of the Floquet Brillouin zone, $\epsilon=\pm\Omega/2$. The Chern numbers of the Floquet bands do not determine the chiral edge mode of these two gaps separately. This gap-resolved information is provided by dynamical winding numbers constructed from the time-evolution operator \cite{Rudner2013,PerezGonzalez2024Anomalous}. Periodically driven topological systems have been realized in photonic lattices and ultracold-atom platforms \cite{Rechtsman2013,Jotzu2014, Dai2025FloquetSSH}.

Weak topological insulators provide a natural framework for studying systems constructed from lower-dimensional topological units. A three-dimensional weak topological insulator can be viewed as a stack of two-dimensional quantum spin Hall layers, with boundary properties that depend on the surface orientation \cite{KaneMele3D2007, Moore3D2007}. Related stacking constructions have been investigated in layered graphene, topological crystalline insulators, and other
quasi-two-dimensional systems \cite{Das2019,Li_2018,Marwa2023, Kim2015,Qian2020,Zou2016,Ion2019,Liu2016,Burkov2011,Basu-02}. In two dimensions, coupled SSH chains offer a platform for examining the relation between the topology of the individual chains and that of the full system.

Periodically driven SSH chains and their extensions exhibit rich quasienergy and edge-state structures \cite{FSSH-01,FSSH-02,FSSH-03, FSSH-04,li2014topological,Agrawal_2022-01, Martinez2026FloquetSSH}. More recently, periodic driving has been shown to induce two-dimensional weak topological phases characterized by two-dimensional Zak phases in extended Qi-Wu-Zhang and altermagnetic Bernevig-Hughes-Zhang models \cite{Zhu2025FloquetWTI}.

The $N$-stacked SSH model introduced in Ref. \cite{Agrawal_2022-02} is a two-dimensional lattice obtained by coupling SSH chains along the transverse direction. It considers two different constructions, depending on whether the SSH chain is in the trivial phase with winding number $w=0$ or in the nontrivial phase with $w=1$. We refer to the construction with topologically trivial constituent SSH chains (\(w=0\) ) as Construction-I, and to the construction with topologically nontrivial constituent chains (\(w=1\) ) as Construction-II. Moreover, in this construction, diagonal interchain bonds generate mixed $k_x-k_y$ dependence in the two-dimensional Bloch Hamiltonian, so that the different bond geometries have distinct Floquet dressing under circular driving. The model, therefore, provides a controlled framework for investigating how the topology of its constituent one-dimensional chains influences the driven two-dimensional phase. It is well studied that circularly polarized light can generate Floquet Chern phases without a static magnetic field \cite{Haldane,CPL-01,CPL-02,CPL-03,CPL-04,CPL-05,CPL-06,CPL-07,FoaTorres2015,Qin2022,Qin2023}. At lower driving frequencies, hybridization between Floquet replicas can also produce Floquet phases with $|C|>1$ \cite{Li_2018}. Here, we apply circular driving to the $N$-stacked SSH model and obtain the driven Hamiltonian in the Floquet-Sambe representation. We characterize the phase diagrams using the Floquet-band Chern number $C_L$ and the directional Zak phases $(Z_x, Z_y)$. Because these band invariants do not identify the topology of the two quasi-energy gaps separately, we further calculate the dynamical winding numbers $W_0$ and $W_{\pi}$, and compare them with quasi-energy spectra in cylindrical geometries. Our central objective is to determine how the topology of the constituent SSH chains affects the driven phase diagram. Compared to Ref. \cite{Agrawal_2022-02}, which examined the topology of the undriven stacked-SSH model, the present work develops the Floquet description of the model under circular driving and compares the driven topologies of the two constituent-chain constructions. 

Optical and dynamical response functions provide complementary probes of light-dressed bands \cite{Huang2021,Karch2010,Hatano2009,Tahir2015,Baykusheva2021}. We therefore also calculate the time-averaged optical Hall conductivity for representative phases of Construction-I and Construction-II. This calculation allows us to examine how the bulk optical response reflects the Floquet-band topology. 

The remainder of the paper is organized as follows. Section \ref{Sec: Model} introduces the static $N$-stacked SSH Hamiltonian and its two constituent-chain constructions. Section \ref{Sec: Driven Hamiltonian} presents the system under circularly polarized driving. The Floquet-Sambe formalism is also introduced here as a tool to investigate driven systems. Section \ref{Results} presents the topological properties of the systems under periodic driving. In Sec. \ref{Sec:OC}, we separately study the role of topological properties on the optical conductivity in the systems. Section \ref{Sec:Experiment} discusses possible experimental realizations of the systems studied in this paper. Finally, we summarize the paper in Sec. \ref{Sec:Discussion}.

\section{\label{Sec: Model}Static Hamiltonian}

We study an $N$-stacked SSH model composed of $N$ coupled SSH chains \cite{Agrawal_2022-02}. In real space, its static Hamiltonian is
\begin{equation}
\begin{split}
H_{N-SSH}&= (1-\eta)\sum_{n_x,n_y} c_{n_x, n_y}^{\dagger A} c_{n_x, n_y}^B + (1+\eta)\sum_{n_x,n_y} c_{n_x+1, n_y}^{\dagger A} c_{n_x, n_y}^B \\
& +\frac{\delta}{2}\, \sum_{n_x,n_y} \bigg[c_{n_x, n_y}^{\dagger A} c_{n_x, n_y+1}^B + c_{n_x, n_y+1}^{\dagger A} c_{n_x, n_y}^B \bigg]\\
& -\frac{\delta}{2}\, \sum_{n_x,n_y} \bigg[c_{n_x+1, n_y}^{\dagger A} c_{n_x, n_y+1}^B + c_{n_x+1, n_y+1}^{\dagger A} c_{n_x, n_y}^B \bigg] + {\rm H. c.}
\end{split}
\label{H_real}
\end{equation}
Here, $\delta$ denotes the interchain hopping amplitude, while $\eta$ controls the dimerization along each SSH chain. In the following, we measure bond lengths in units of the lattice constant $a$. Moreover, $A$ and $B$ denote two internal orbitals centered at the same Bravais-lattice site. This co-located orbital embedding corresponds to the SSH$_0$-type construction discussed in Ref. \cite{FuchsPiechon2021}.  To examine the energy spectrum
and topological properties, we write the static Hamiltonian in quasimomentum ($k$-) space as
\begin{equation}
H_{N-SSH}(\bs{k}) = \bs{h} \bs{\cdot} \bs{\sigma}
\label{Eq-NSSH}
\end{equation}
\begin{align*}
h_x(\textbf{k}) =& [(1+\cos k_x) + (1-\cos k_x) \, (\delta \cos k_y - \eta)] \\
h_y(\textbf{k}) =& [(1+\eta) - \delta \cos k_y] \, \sin k_x 
\end{align*}
The explicit symmetry operators and the corresponding symmetry classification of the undriven Hamiltonian are presented in Appendix~\ref{app:static_symmetry}. The $N$ stacked SSH model was introduced and studied in detail in Ref. \cite{Agrawal_2022-02}. Here, we consider the translationally invariant limit, in which the stacking direction is also treated as a periodic spatial direction, and the system admits a two-dimensional Bloch description in $(k_x, k_y)$. The real-space geometry and the two constructions of the model are as defined in Ref. \cite{Agrawal_2022-02}.

For a fixed transverse momentum $k_y$,  Eq. \eqref{Eq-NSSH} can be written as,
\begin{align}
h_x(\mathbf{k})&=t_1(k_y)+t_2(k_y)\cos k_x,\\
h_y(\mathbf{k})&=t_2(k_y)\sin k_x,
\end{align}
which is an effective SSH Hamiltonian. Here,
\begin{equation}
t_1(k_y)=1-\eta+\delta\cos k_y,
\qquad
t_2(k_y)=1+\eta-\delta\cos k_y.
\label{Eq-t1t2}
\end{equation}
For $a=x,y$, let $k_\perp$ denote the transverse momentum, with $k_\perp=k_y$ for $a=x$ and $k_\perp=k_x$ for $a=y$. For each gapped fixed-$k_\perp$ slice, we define the normalized hybrid Zak phase of the lower static band as
\begin{equation}
Z_a(k_\perp)
=
\frac{\mathrm{i}}{2\pi}
\oint_{\mathrm{BZ}_a}
\left\langle u_-(\mathbf{k})\middle|
\partial_{k_a}u_-(\mathbf{k})
\right\rangle\,dk_a
\pmod{1},
~~a=x,\, y.
\label{eq:static-hybrid-zak}
\end{equation}
Here, $|u_-(\mathbf{k})\rangle$ denotes the lower-band eigenstate of the undriven Hamiltonian. With the Bloch basis and unit-cell convention adopted in  Eq. (\ref{H_real}), the condition $|t_2(k_y)|>|t_1(k_y)|$ becomes $\eta-\delta\cos k_y>0$, and hence
\begin{equation}
Z_x(k_y)=
\begin{cases}
\dfrac{1}{2}, & \eta-\delta\cos k_y>0,\\[1mm]
0, & \eta-\delta\cos k_y<0.
\end{cases}
\label{eq:static-Zx}
\end{equation}
At $(\eta-\delta\cos k_y)=0$, the slice is gapless, so $Z_x(k_y)$ is not defined.
However, for every gapped fixed-$k_x$ loop, $Z_y(k_x)=0$ because the Bloch vector retraces its path as $k_y$ traverses the Brillouin zone. For the parameters $\delta=1$ and $\eta=\pm0.5$ considered below, the fixed-$k_x=\pi$ loop crosses the Dirac points, so its Zak phase is undefined. 

For $\delta=1$ and $\eta=\pm0.5$, the condition $\delta\cos k_y=\eta$ gives two symmetry-related Dirac band-touching points at $k_x=\pi$. Changing $\eta\rightarrow-\eta$ together with $k_y\rightarrow k_y+\pi$ exchanges $t_1$ and $t_2$; hence, the two undriven periodic bulk Hamiltonians are unitarily equivalent and have identical spectra. In the decoupled-chain limit (\(\delta=0\)), Construction-I corresponds to \(\eta=-0.5\) and \(w=0\), whereas Construction-II corresponds to \(\eta=+0.5\) and \(w=1\). For the parameter set used throughout this work, the undriven system is gapless, with Dirac points in the bulk spectrum. Consequently, the topological phases discussed below, in the presence of circularly polarized driving, are genuinely drive-induced Floquet phases rather than adiabatic continuations of distinct static insulating phases. 

\section{\label{Sec: Driven Hamiltonian}Driven Hamiltonian} 

We study the effect of periodic driving by applying a time-periodic vector potential of the form
\begin{equation}
\bs{A}_{\chi}(t)
=\left(A_{0}\cos\Omega t,\,
\chi A_{0}\sin\Omega t\right),
\qquad \chi=\pm1,
\label{Eq-vector-potential}
\end{equation}
where $\chi$ specifies the helicity of the drive. Throughout this work, we choose $\chi=-1$. 
This driving is induced in the system by the Peierls substitution, which modifies the form of the quasi-momenta $k_x$ and $k_y$ as 
\[k_x(t) \rightarrow k_x + A_x(t)\, ; \, k_y(t) \rightarrow k_y + A_y(t) \]\\
Because the two internal orbitals share the same real-space center, the local coupling $t_0(1-\eta)$ has zero bond displacement and therefore does not acquire a Peierls phase. The field then enters only through Peierls phases on the intersite hopping.

The time-periodic Hamiltonian in the $\bs{k}$-space is
\begin{equation}
\begin{split}
H_{N-SSH}(\textbf{k},t) =& [(1+\cos k_x(t)) + (1-\cos k_x(t)) \\
& \times (\delta \cos k_y(t) - \eta)] \sigma_x \\
&+ [(1+\eta) - \delta \cos k_y(t)] \, \sin k_x(t) \, \sigma_y
\end{split}
\label{Eq-Driven-NSSH}
\end{equation}
We also set $\hbar=1$, so that the driving frequency $\Omega$ is expressed in units of $t_0$ and $A_0$ is the dimensionless Peierls amplitude. To solve the time-periodic equation, we use the Floquet replica method and compute the effective Hamiltonian as an infinite-dimensional matrix in the frequency space.

\subsection{Floquet theory}

For a time-periodic Hamiltonian, $H(t+T)=H(t)$, Floquet's theorem gives \cite{Floquettheory1883,shirley1965solution,Eckardt_2015}
\begin{equation}
|\psi_n(t)\rangle=e^{-i\epsilon_n t}|u_n(t)\rangle,
\qquad
|u_n(t+T)\rangle=|u_n(t)\rangle,
\end{equation}
where $\epsilon_n$ is the quasienergy and $|u_n(t)\rangle$ is the corresponding Floquet mode. The Floquet modes satisfy
\begin{equation}
\left[H(t)-i\frac{\partial}{\partial t}\right]
|u_n(t)\rangle
=
\epsilon_n|u_n(t)\rangle.
\label{eq:TPSE}
\end{equation}
Equivalently, they are eigenstates of the evolution operator over one driving period \cite{JNBandTGS},
\begin{equation}
\hat{U}(t_0+T,t_0)|u_n(t_0)\rangle
=
e^{-i\epsilon_nT}|u_n(t_0)\rangle.
\end{equation}
The Hamiltonian and Floquet modes can be expanded as
\[
H(t)=\sum_m e^{-im\Omega t}H^{(m)},
\qquad
|u_n(t)\rangle
=
\sum_m e^{-im\Omega t}|u_n^{(m)}\rangle ,
\]
where $m=0,\pm1,\pm2,\ldots$ labels the photon sectors in the extended Floquet space. Quasienergies are defined modulo $\Omega$, and we choose their representatives in the first Floquet Brillouin zone, $-\Omega/2<\epsilon\leq\Omega/2$. The Fourier components are
\begin{equation}
H^{(m)}(\mathbf{k})
=
\frac{1}{T}\int_0^T
H(\mathbf{k},t)e^{im\Omega t}\,dt.
\label{Eq-Fourier-def}
\end{equation}
Substitution into  Eq. \eqref{eq:TPSE} gives the Floquet--Sambe eigenvalue equation
\begin{equation}
\epsilon_n|u_n^{(m)}\rangle
=
\sum_{m'}
\left[
H^{(m-m')}-m\Omega\delta_{mm'}
\right]
|u_n^{(m')}\rangle .
\end{equation}
This equation is defined in the extended Hilbert space $\mathcal{H}\otimes\mathcal{T}$ \cite{Sambe}, where $H^{(m-m')}$ couples photon sectors $m$ and $m'$. The infinite Floquet-Sambe matrix is truncated numerically, and convergence is checked by increasing the number of photon sectors. Lower driving frequencies generally require more sectors because more Floquet replicas participate in the hybridization.

\subsection{Periodically driven ${\bs N}$ stacked SSH model: Floquet formalism}
\label{subsec:driven_formalism}

For any m ($m\in\mathbb{Z}$), the Fourier component of the driven Hamiltonian can be written as
\begin{equation}
H_{N\text{-}SSH}^{(m)}(\bs{k})
=d_x^{(m)}(\bs{k})\sigma_x+d_y^{(m)}(\bs{k})\sigma_y,
\qquad d_z^{(m)}=0.
\label{Eq-Hm-general}
\end{equation}
It is useful to define
\begin{equation}
\alpha_m=\frac{m\pi}{2},\qquad
\beta_m=\frac{m\pi}{4},\qquad
K_m=k_x+\alpha_m.
\label{Eq-harmonic-phases}
\end{equation}
The two nonzero components are
\begin{subequations}
\label{Eq-F-comp}
\begin{align}
d_x^{(m)}={}&(1-\eta)\delta^{\mathrm K}_{m0}
 +(1+\eta)J_m(A_0)\cos K_m
 \nonumber\\
&+\delta e^{-i\alpha_m}J_m(A_0)\cos(k_y+\alpha_m)
 \nonumber\\
&-\delta J_m(\sqrt{2}A_0)
 \bigl[\cos\beta_m\cos K_m\cos k_y
 \nonumber\\
&\hspace{29mm}
 +i\sin\beta_m\sin K_m\sin k_y\bigr],
\label{Eq-F-comp-x}\\[1mm]
d_y^{(m)}={}&(1+\eta)J_m(A_0)\sin K_m
 \nonumber\\
&-\delta J_m(\sqrt{2}A_0)
 \bigl[\cos\beta_m\sin K_m\cos k_y
 \nonumber\\
&\hspace{29mm}
 -i\sin\beta_m\cos K_m\sin k_y\bigr].
\label{Eq-F-comp-y}
\end{align}
\end{subequations}
Here, $J_m$ denotes the Bessel function of the first kind, and $\delta^{\mathrm K}_{m0}$ is the Kr\"onecker delta, distinguished from the interchain hopping $\delta$. Under the Peierls substitution, the momentum-independent intracell term remains unchanged, whereas the momentum-dependent hopping terms acquire the bond-dependent Bessel factors $J_m(A_0)$ and $J_m(\sqrt{2}A_0)$.
These Fourier components obey the Hermiticity condition
\begin{equation}
H_{N\text{-}SSH}^{(-m)}(\bs{k})
=\left[H_{N\text{-}SSH}^{(m)}(\bs{k})\right]^{\dagger},
\label{Eq-Hm-hermiticity}
\end{equation}
and reduce to the static Hamiltonian in the undriven limit,
\begin{equation}
\lim_{A_0\rightarrow0}H_{N\text{-}SSH}^{(m)}(\bs{k})
=\delta^{\mathrm K}_{m0}H_{N\text{-}SSH}(\bs{k}).
\label{Eq-Hm-static-limit}
\end{equation}
For the numerical calculations below, the photon index is truncated to $m=-4,\ldots,+4$, giving nine photon sectors. Diagonalizing the resulting Floquet-Sambe Hamiltonian at each $\bs{k}$ yields the quasienergies and Floquet modes used to construct the phase diagrams. We explicitly tested the convergence of this truncation at $\Omega=2.0$, the lowest driving frequency used in the phase diagrams, by increasing the number of photon sectors to eleven and thirteen. The calculated Floquet-band Chern number $C_L$ remained unchanged.

\section{Floquet topological phases}
\label{Results}

In the following sections, we compare the Floquet topology induced by circular driving in Construction-I and Construction-II. After presenting their bulk phase diagrams, we examine the origin of their distinct phase structures and characterize the gap-resolved edge states using the dynamical winding numbers. 

\subsection{Construction-dependent phase diagrams }

\begin{figure}[b]
\includegraphics[width=0.5\textwidth]{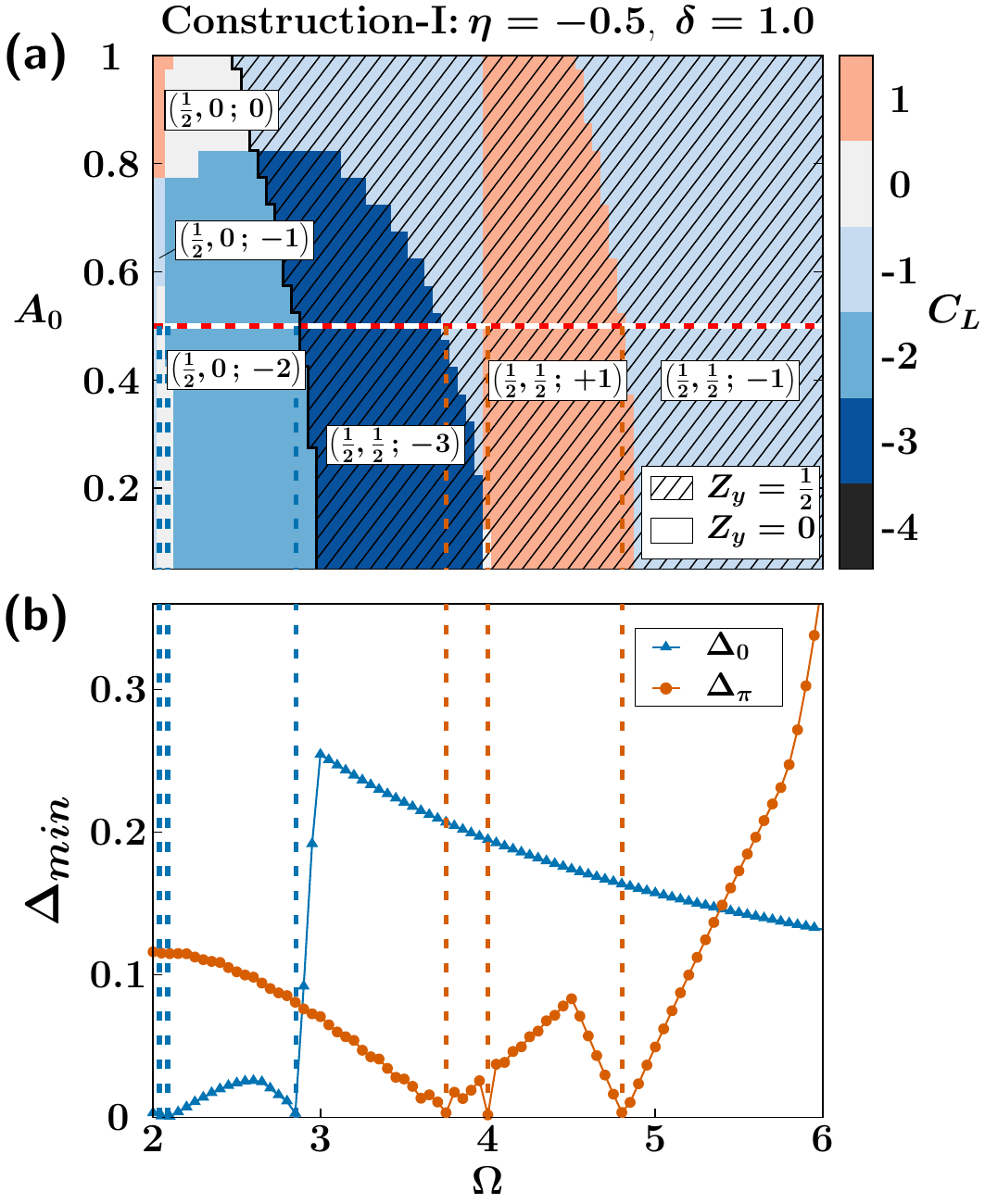}
\caption{Floquet phase diagram in the $(A_0,\Omega)$ plane for $\eta=-0.5$ and $\delta=1$, corresponding to Construction-I. The color scale represents $C_L$, while the hatched and unhatched regions correspond to $Z_y=1/2$ and $Z_y=0$, respectively. The labels give $(Z_x,Z_y;C_L)$. Panel (b) shows the minimum bulk quasienergy gaps $\Delta_0$ and $\Delta_\pi$ at $\epsilon=0$ and $\epsilon=\pm\Omega/2$, respectively, as functions of $\Omega$ along the $A_0=0.5$ cut marked by the red horizontal line in panel (a). The blue and orange vertical dashed lines mark near-zero minima of the corresponding gaps.}
\label{Fig-02}
\end{figure}

\begin{figure}
\includegraphics[width=0.5\textwidth]{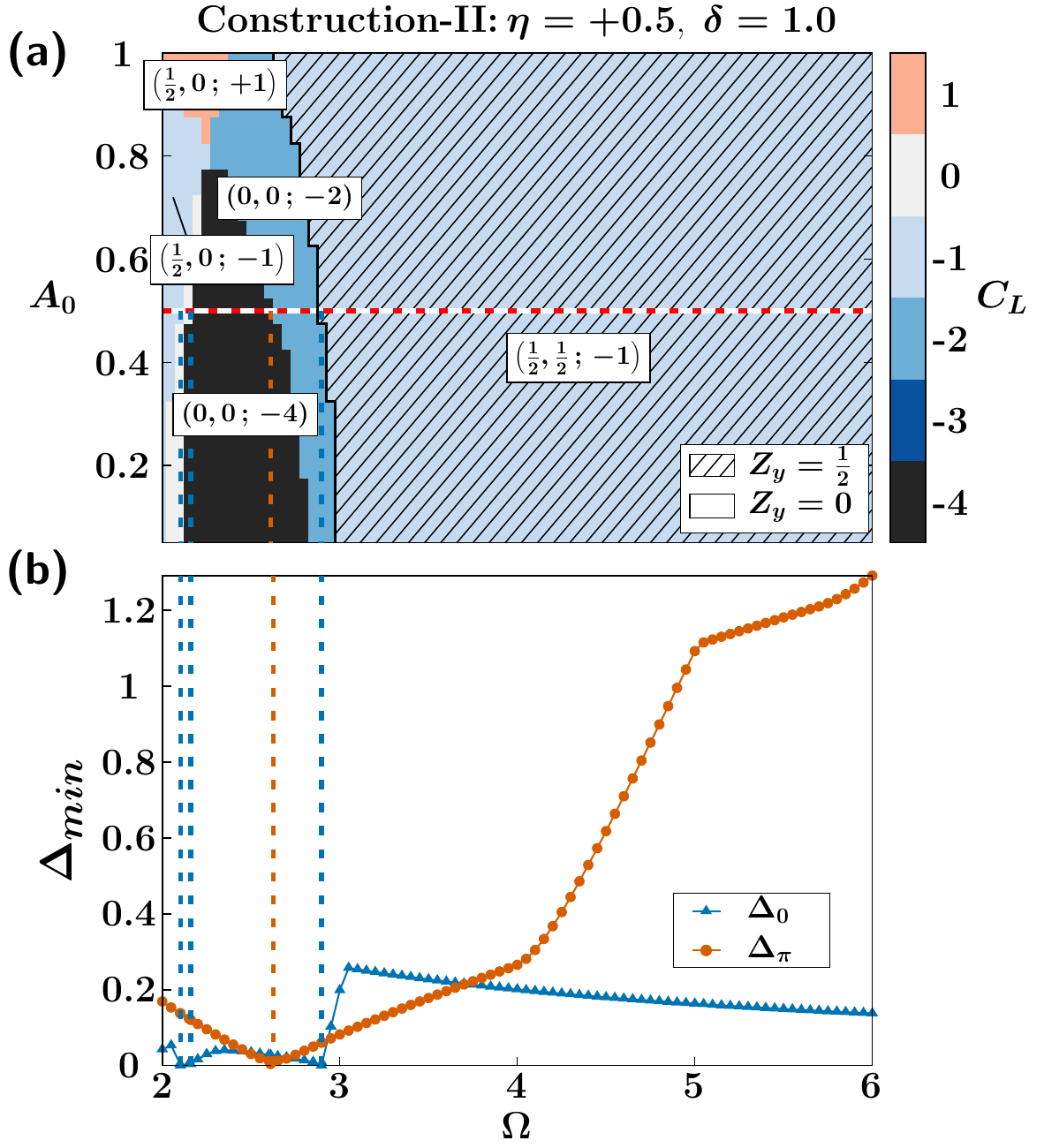}
\caption{Floquet phase diagram for $\eta=+0.5$ and $\delta=1$, corresponding to Construction-II. The notation is the same as in  Fig. \ref{Fig-02}. Panel (b) shows $\Delta_0$ and $\Delta_\pi$ as functions of $\Omega$ along the $A_0=0.5$ cut marked in panel (a). The blue and orange vertical dashed lines mark near-zero minima of the corresponding gaps.}
\label{Fig-04}
\end{figure}

We characterize each gapped Floquet phase in the $(A_0,\Omega)$ plane by the triplet
\begin{equation}
    (Z_x,Z_y;C_L),
    \label{eq:phase-label}
\end{equation}
where $C_L$ is the Chern number of the lower Floquet band in the first Floquet Brillouin zone, $-\Omega/2<\epsilon<0$, and $Z_x$ and $Z_y$ are its directional Zak phases. The directional Zak phases $Z_x$ and $Z_y$ characterize the Floquet states along selected one-dimensional loops in momentum space. They provide additional information beyond the Chern number and can distinguish Floquet phases with the same Chern number. Unlike the Chern number, the Zak phases are defined along one-dimensional loops, and their absolute values depend on the choice of Bloch basis and unit cell.

Let $|u_{L\mathbf{k}}(t)\rangle$ denote the periodic Floquet mode of the isolated lower band. Its time-averaged Berry connection is
\begin{equation}
    \mathcal{A}_a(\mathbf{k})
    =
    i\Bigl\langle\Bigl\langle
    u_{L\mathbf{k}}
    \Bigr|\Bigl.
    \partial_{k_a}u_{L\mathbf{k}}
    \Bigr\rangle\Bigr\rangle,
    \qquad a=x,y,
    \label{eq:zak-connection}
\end{equation}
where the double brackets denote the time-averaged Sambe inner product. The corresponding Berry curvature and Chern number \cite{Fukui2005} are
\begin{equation}
    \mathcal{F}_{xy}(\mathbf{k})
    =
    \partial_{k_y}\mathcal{A}_x
    -
    \partial_{k_x}\mathcal{A}_y,
    \label{eq:berry-curvature}
\end{equation}
and
\begin{equation}
    C_L
    =
    \frac{1}{2\pi}
    \int_{\mathrm{BZ}}
    d^2k\,\mathcal{F}_{xy}(\mathbf{k}).
    \label{eq:chern-number}
\end{equation}
Although circular driving breaks the time-reversal symmetry, the driven Hamiltonian retains a dynamical inversion symmetry. Within the Bloch basis adopted here, the undriven Hamiltonian satisfies
\[\sigma_x H_{N\text{-}SSH}(\mathbf{k})\sigma_x = H_{N\text{-}SSH}(-\mathbf{k}).\]
Moreover, since $\mathbf{A}(t+T/2)=-\mathbf{A}(t)$, the driven Hamiltonian obeys
\[\sigma_x H_{N\text{-}SSH}(\mathbf{k},t)\sigma_x = H_{N\text{-}SSH}(-\mathbf{k},t+T/2).\]
Defining $\widetilde{\mathcal I}=\sigma_x\tau_{T/2}$, where $\tau_{T/2}f(t)=f(t+T/2)$, the Floquet-Sambe operator $\mathcal{Q}(\mathbf{k})=H_{N\text{-}SSH}(\mathbf{k},t)-i\partial_t$ therefore satisfies
\[\widetilde{\mathcal I}\mathcal{Q}(\mathbf{k}) \widetilde{\mathcal I}^{-1} = \mathcal{Q}(-\mathbf{k}), \qquad \widetilde{\mathcal I}^{\,2}=1.\]
This operation is a twofold time-screw symmetry, consisting of a twofold spatial rotation and a translation by half a driving period \cite{Morimoto2017TimeGlide}. In the Fourier basis $e^{-im\Omega t}$, its Sambe-space representation is
$(\widetilde{\mathcal I})_{mm'}=(-1)^m\sigma_x\delta_{mm'}$.

The momentum-space lines $k_y=\pi$ and $k_x=\pi$ are invariant under this symmetry, since $-\pi$ and $\pi$ are equivalent modulo a reciprocal lattice vector. Along these loops, the Zak phases are quantized to $0$ or $1/2$ modulo one, provided the lower Floquet band remains isolated. We therefore define the fixed-line Zak phases of the lower Floquet band as
\begin{equation}
Z_a=
\left.
\frac{1}{2\pi}\oint_{\mathrm{BZ}_a}
\mathcal{A}_a(\mathbf{k})\,dk_a
\right|_{k_\perp=\pi}
\pmod{1},
\qquad a=x,y.
\label{eq:fixed-line-zak}
\end{equation}

Provided that the lower Floquet band remains isolated along the corresponding loop, dynamical inversion quantizes each Zak phase to $0$ or $1/2$ modulo one \cite{Hughes2011Inversion}. These values are quoted using the Bloch basis, unit cell, and real-space origin adopted in Eq. (\ref{H_real}), since the absolute Zak phase depends on these conventions \cite{Rhim2017Zak}. Figures \ref{Fig-02} and \ref{Fig-04} show the resulting phase diagrams for $\eta=-0.5$ and $\eta=+0.5$, respectively, with $\delta=1$.

The Zak sectors differ between the two constructions. For $\eta=-0.5$, Fig. \ref{Fig-02} has $Z_x=1/2$ throughout the displayed parameter range. For $\eta=+0.5$,  Fig. \ref{Fig-04} contains regions with both $Z_x=0$ and $Z_x=1/2$. In both diagrams, the hatched and unhatched regions distinguish $Z_y=1/2$ and $Z_y=0$, respectively. The Zak phases can therefore distinguish Floquet topological phases that share the same Chern number.

Construction-I supports phases with $C_L=0,\pm1,-2$, and $-3$, whereas Construction-II supports $C_L=0,\pm1,-2$, and $-4$. In both cases, phases with $|C_L|>1$ occur predominantly in the low-frequency regime, while the higher-frequency regions are dominated by $|C_L|=1$. The phase boundaries depend on both $A_0$ and $\Omega$. Their predominantly horizontal structure indicates a stronger dependence on the driving frequency than on the amplitude over the amplitude range considered here. The two constructions, therefore, display different Chern-number patterns and Zak phase sectors. For the two quasienergy bands
$\epsilon_-(\mathbf{k})<0<\epsilon_+(\mathbf{k})$ in the first Floquet Brillouin zone, the minimum gaps are
\begin{align*}
\Delta_0 &= \min_{\mathbf{k}}
\left[\epsilon_+(\mathbf{k})-\epsilon_-(\mathbf{k})\right],\\
\Delta_\pi &= \min_{\mathbf{k}}
\left[\Omega-\epsilon_+(\mathbf{k})+\epsilon_-(\mathbf{k})\right].
\end{align*}
The changes in $C_L$, $W_0$, or $W_\pi$ occur when the corresponding bulk quasienergy gap closes and reopens. For Construction-I, Fig. \ref{Fig-02}(b) follows the $A_0=0.5$ cut in Fig. \ref{Fig-02}(a). The zero-quasienergy gap $\Delta_0$ has near-zero minima between $\Omega\simeq2.0$ and $2.1$ and near $\Omega\simeq2.85$, whereas the Floquet-zone-edge gap $\Delta_\pi$ has minima near $\Omega\simeq3.75$, $4.0$, and $4.8$. The corresponding colored dashed lines locate these frequencies on the phase diagram. For Construction-II, Fig. \ref{Fig-04}(b) shows the same frequency scan. The minimum of $\Delta_\pi$ near $\Omega\simeq2.6$ lies between the representative $C_L=-4$ phase at $\Omega=2.5$ and the $C_L=-2$ phase at $\Omega=2.8$. The zero-quasienergy gap $\Delta_0$ has separate near-zero minima around $\Omega\simeq2.1$ and $2.9$. The changes in $C_L$ across these boundaries coincide with closings and reopenings of the corresponding Floquet quasienergy gaps, confirming that the distinct Chern sectors are separated by bulk Floquet gap closings. These gap curves identify which quasienergy gap becomes small along the cut; a minimum alone does not establish the change in a topological invariant. The physical origin of these differences is examined in the following subsection.

\subsection{Origin of the distinct phase structures}

Although the two static constructions are related by a unitary transformation that exchanges the effective intracell and intercell hopping amplitudes, this equivalence is not preserved after minimal coupling to the time-dependent vector potential. The reason is that the exchanged hopping processes correspond to different physical bond vectors and therefore acquire different time-dependent Peierls phases under circular driving. This is evident from the zero-harmonic component:
\begin{equation}
\begin{aligned}
&t_{1}^{(0)}(k_y+\pi, -\eta)
=(1+\eta)-\delta J_0(A_0)\cos k_y,~{\rm and}\\
&t_{2}^{(0)}(k_y, \eta)
=(1+\eta)J_0(A_0)
-\delta J_0(\sqrt{2}A_0)\cos k_y .
\end{aligned}
\label{Eq-driven-hopping-comparison}
\end{equation}
These expressions coincide in the undriven limit $A_0=0$, but are generally different for finite $A_0$. Thus, the zero-harmonic component illustrates how the bond-geometry-dependent Floquet dressing distinguishes the two constructions. In the low-frequency regime, however, the complete phase structure is determined by all Fourier harmonics and their inter-replica hybridization.

In the undriven system, $k_y$ enters only through $\cos k_y$, so the Bloch vector retraces its path and $Z_y=0$ for every gapped fixed-$k_x$ loop. The drive-induced $\sin k_y$ terms in  Eq. \eqref{Eq-F-comp} remove this retracing constraint, allowing gapped driven phases with $Z_y=1/2$ as well as $Z_y=0$. Although the coefficients of these terms are independent of $\eta$, the value of $Z_y$ depends on the full Floquet Hamiltonian, including the $\eta$-dependent hopping terms and their inter-replica hybridization.

For $Z_x$, the relevant loop is $k_y=\pi$, where the explicit $\sin k_y$ terms vanish. Both constructions have $Z_x=1/2$ for the lower static band along this loop. Under driving, the construction-dependent Peierls dressing of the remaining hopping terms and their inter-replica hybridization determine the Floquet eigenstates. Dynamical inversion symmetry restricts $Z_x$ to either $0$ or $1/2$, provided the lower Floquet band is isolated along the loop. A transition between these values requires a gap to close in this loop. For $\eta=-0.5$, the calculated $Z_x$ is $1/2$ throughout the gapped regions displayed in  Fig. \ref{Fig-02}, whereas for $\eta=+0.5$, both $Z_x=0$ and $Z_x=1/2$ occur in  Fig. \ref{Fig-04}. Thus, the construction dependence is not limited to the value of the Floquet-band Chern number; it can also determine which quasienergy gap carries the corresponding chiral edge modes.

\subsection{Demonstration of the anomalous edge states}

\begin{figure}[t]
\includegraphics[width = 0.45\textwidth]{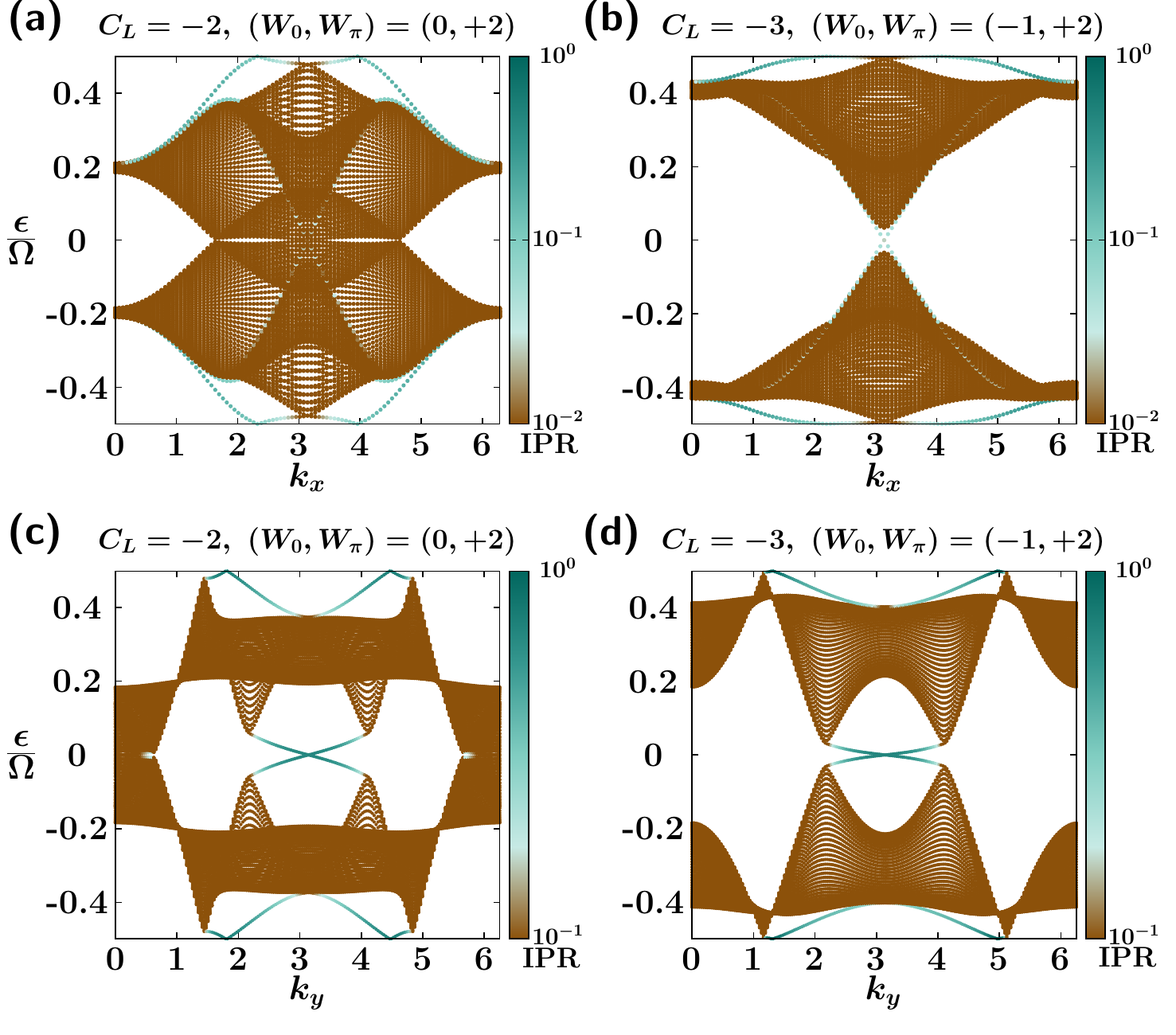}
\caption{Quasienergy spectra in cylindrical geometry for Construction-I. Panels (a) and (c) show $(A_0,\Omega)=(0.5,2.5)$, for which $C_L=-2$ and $(W_0,W_{\pi})=(0,+2)$. Panels (b) and (d) show $(A_0,\Omega)=(0.5,3.5)$, for which $C_L=-3$ and $(W_0,W_{\pi})=(-1,+2)$. Panels (a) and (b) are periodic along $x$ and open along $y$, whereas panels (c) and (d) are periodic along $y$ and open along $x$. The color scale represents the inverse participation ratio (IPR), which highlights the edge-localized states.}
\label{Fig-03}
\end{figure}

\begin{figure}
\includegraphics[width = 0.45\textwidth]{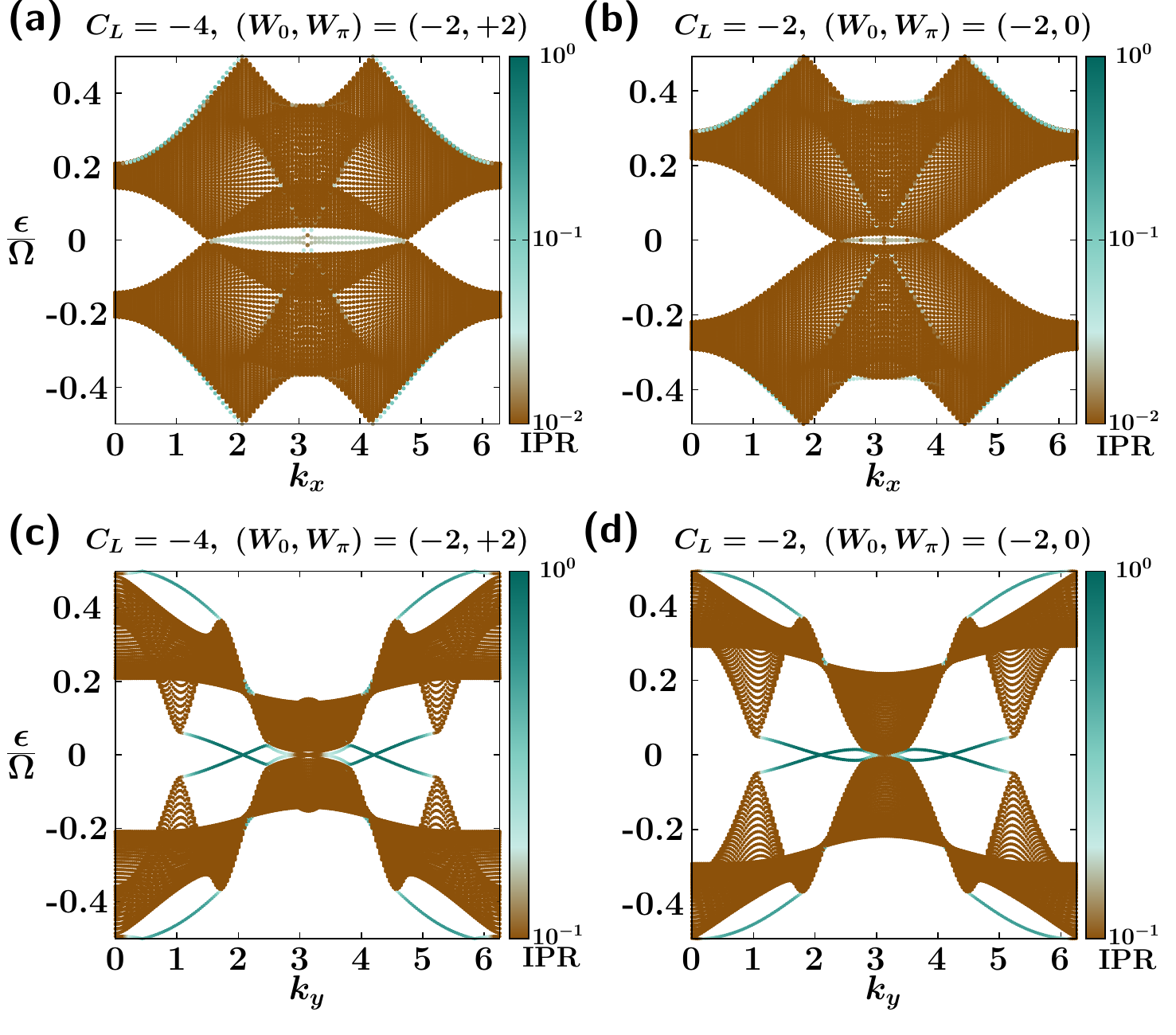}
\caption{Quasienergy spectra in cylindrical geometry for Construction-II.  Panels (a) and (c) show $(A_0,\Omega)=(0.5,2.5)$, for which $C_L=-4$ and $(W_0,W_{\pi})=(-2,+2)$. Panels (b) and (d) show $(A_0,\Omega)=(0.5,2.8)$, for which $C_L=-2$ and $(W_0,W_{\pi})=(-2,0)$. Panels (a) and (b) are periodic along $x$ and open along $y$, whereas panels (c) and (d) are periodic along $y$ and open along $x$. The color scale represents the inverse participation ratio (IPR), which highlights the edge-localized states.}
\label{Fig-05}
\end{figure}

The periodicity of quasienergy modulo $\Omega$ distinguishes Floquet bulk--boundary correspondence from its static counterpart. In the first Floquet Brillouin zone, $-\Omega/2<\epsilon\leq\Omega/2$, edge states may traverse both the central gap at $\epsilon=0$ and the zone-edge gap at $\epsilon=\pm\Omega/2$. The Chern numbers of the Floquet bands alone do not specify the net chirality in each of these two gaps. We therefore use the gap-resolved dynamical winding number of the full time evolution \cite{Rudner2013}.

For a driving period $T=2\pi/\Omega$, the bulk evolution operator is
\begin{equation}
 U(\mathbf{k},t)=\mathcal{T}
 \exp\left[-i\int_{0}^{t}H(\mathbf{k},t')\,dt'\right].
 \label{eq:bulk-time-evolution}
\end{equation}
For a quasienergy $\varepsilon$ lying in an open bulk gap, we define the gap-dependent effective Hamiltonian and the corresponding periodized evolution as
\begin{align}
 H_{\mathrm{eff}}^{(\varepsilon)}(\mathbf{k})
 &=\frac{i}{T}\log_{\varepsilon}U(\mathbf{k},T),
 \label{eq:gap-effective-hamiltonian}\\
 U_{\varepsilon}(\mathbf{k},t)
 &=U(\mathbf{k},t)
 e^{iH_{\mathrm{eff}}^{(\varepsilon)}(\mathbf{k})t}.
 \label{eq:periodized-evolution}
\end{align}
Here, $\log_{\varepsilon}$ has its branch cut through $e^{-i\varepsilon T}$. Since $e^{-iH_{\mathrm{eff}}^{(\varepsilon)}T}=U(\mathbf{k},T)$, the periodized evolution satisfies
$U_{\varepsilon}(\mathbf{k},0)=U_{\varepsilon}(\mathbf{k},T)=\unit$. Defining
$Q_{\mu}^{(\varepsilon)}=U_{\varepsilon}^{-1}\partial_{\mu}U_{\varepsilon}$ for $\mu=t,k_x,k_y$, the dynamical winding number is
\begin{equation}
 W_{\varepsilon}=\frac{1}{8\pi^{2}}
 \int_{0}^{T}dt\int_{\mathrm{BZ}}d^{2}\mathbf{k}\,
 \operatorname{Tr}\left\{
 Q_{t}^{(\varepsilon)}
 \left[Q_{k_x}^{(\varepsilon)},Q_{k_y}^{(\varepsilon)}\right]
 \right\}.
 \label{eq:dynamical-winding}
\end{equation} 
We denote the winding numbers of the central and Floquet-zone-edge gaps by $W_0\equiv W_{\varepsilon=0}$ and $W_{\pi}\equiv W_{\varepsilon=\Omega/2}$, respectively; the subscript $\pi$ refers to the Floquet eigenphase $\varepsilon T=\pi$. With the convention used here, the total Chern number $C_L$ of the Floquet bands in $-\Omega/2<\epsilon<0$ obeys
\begin{equation}
 C_L=W_0-W_{\pi}.
 \label{eq:chern-winding-relation}
\end{equation}
A nonzero $W_{\pi}$ identifies chiral edge modes at the Floquet-zone boundary. We use ``anomalous'' in this gap-resolved sense, which does not require the Floquet-band Chern number to vanish. Reversing the drive helicity is equivalent to $k_y\rightarrow-k_y$ and reverses the signs of $C_L$, $W_0$, and $W_{\pi}$, while leaving the quasienergy gaps and phase boundaries unchanged.

Figures \ref{Fig-03} and \ref{Fig-05} show cylindrical quasienergy spectra for the two constructions. For Construction-I, $(A_0,\Omega)=(0.5,2.5)$ gives $(W_0,W_{\pi})=(0,+2)$ and $C_L=-2$. The central gap, therefore, has zero net chirality, whereas the zone-edge gap has net chirality $W_{\pi}=+2$, corresponding to two co-propagating modes on a given boundary. At $(A_0,\Omega)=(0.5,3.5)$, $(W_0,W_{\pi})=(-1,+2)$ and $C_L=-3$, so both gaps are topologically active. For Construction-II, $(A_0,\Omega)=(0.5,2.5)$ gives $(W_0,W_{\pi})=(-2,+2)$ and $C_L=-4$, with nonzero net chirality in both gaps. At $(A_0,\Omega)=(0.5,2.8)$, $(W_0, W_{\pi})=(-2,0)$ and $C_L=-2$, so the net chirality is confined to the central gap. The IPR-resolved spectra for both boundary orientations exhibit edge-localized branches consistent with these gap winding numbers. 

The two $C_L=-2$ phases therefore have inequivalent gap topology: $(W_0,W_{\pi})=(0,+2)$ for Construction-I and $(-2,0)$ for
Construction-II. This distinction is also obtained at identical driving parameters. At $(A_0,\Omega)=(0.7,2.6)$, both constructions have $C_L=-2$, but their winding numbers remain $(0,+2)$ and $(-2,0)$, respectively. Thus, the constituent-chain construction changes how the same Floquet-band Chern number is distributed between the central and zone-edge gaps.

\section{\label{Sec:OC} Optical conductivity in irradiated $N$ stacked SSH model}

\begin{figure}
\includegraphics[width = 0.5\textwidth]{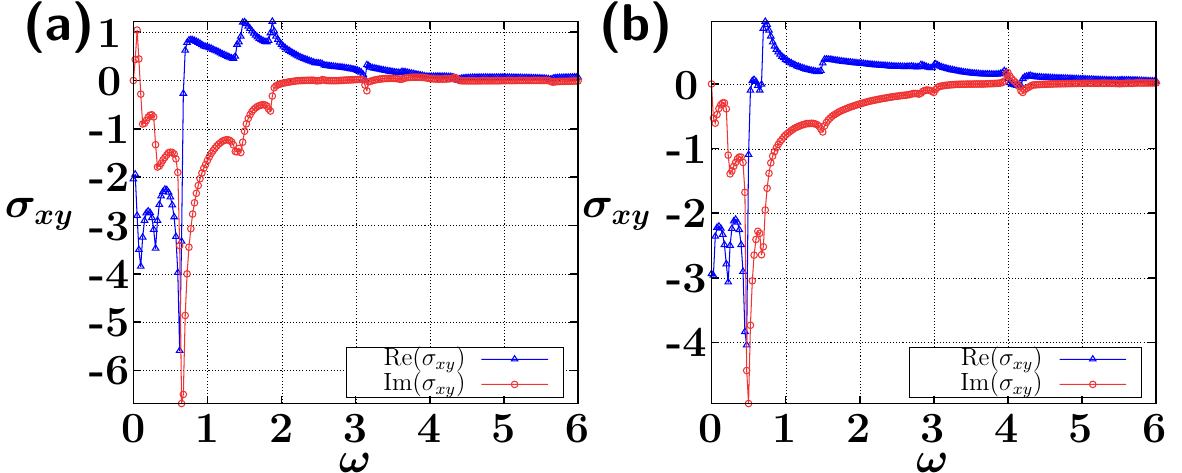}
\caption{Real and imaginary parts of the optical Hall conductivity for Construction-I at $A_0=0.5$. Panels (a) and (b) correspond to $\Omega=2.5$, with $C_L=-2$ and $(W_0,W_\pi)=(0,+2)$, and $\Omega=3.5$, with
$C_L=-3$ and $(W_0,W_\pi)=(-1,+2)$, respectively. The conductivity is expressed in units of $e^2/h$. }
\label{OC-Fig-05}
\end{figure}

\begin{figure}
\includegraphics[width = 0.5\textwidth]{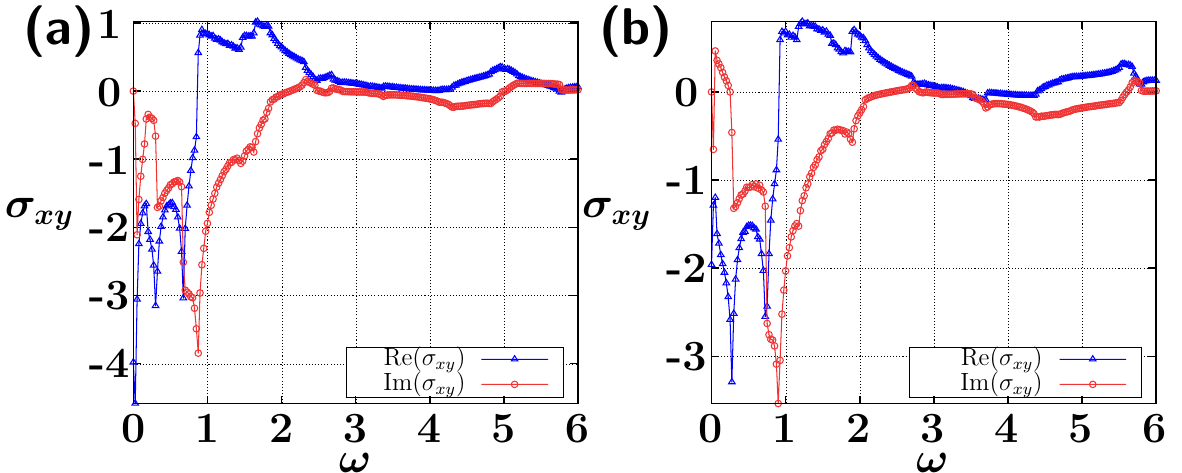}
\caption{Real and imaginary parts of the optical Hall conductivity for Construction-II at $A_0=0.5$. Panels (a) and (b) correspond to $\Omega=2.5$, with $C_L=-4$ and $(W_0,W_\pi)=(-2,+2)$, and $\Omega=2.8$, with $C_L=-2$ and $(W_0,W_\pi)=(-2,0)$, respectively. The conductivity is expressed in units of $e^2/h$. }
\label{OC-Fig-4d}
\end{figure}

We next examine the bulk optical response of the representative Floquet phases discussed above. Since the Floquet modes are periodic in time, optical transitions can be accompanied by the absorption or emission of an integer number of drive photons. We calculate the time-averaged conductivity using the Floquet-Kubo formalism \cite{Vega2020,DabiriCheraghchiSadeghi2022}. Let $\epsilon_{\alpha\mathbf{k}}$ and $\lvert u_{\alpha\mathbf{k}}(t)\rangle$ denote a quasienergy and its corresponding $T$-periodic Floquet mode, respectively.  We introduce
\begin{equation}
 \Delta_{\alpha\gamma}^{(m)}(\mathbf{k})
 = \epsilon_{\alpha\mathbf{k}}-\epsilon_{\gamma\mathbf{k}}-m\Omega,
 \label{eq:floquet-transition-energy}
\end{equation}
where $m\in\mathbb{Z}$ labels an $m$-photon process.  At zero temperature, the time-averaged optical conductivity is
\begin{align}
 \sigma^{ab}(\omega)
 ={}&-\frac{2\pi i}{N_{\mathbf{k}}}
 \sum_{\mathbf{k}}
 \sum_{\alpha,\,\gamma>\alpha}
 \sum_m
 \frac{f_{\alpha\mathbf{k}}-f_{\gamma\mathbf{k}}}
      {\Delta_{\alpha\gamma}^{(m)}(\mathbf{k})}
 \nonumber\\
 &\times\left[
 \frac{j_{\alpha\gamma}^{b(m)}j_{\gamma\alpha}^{a(-m)}}
      {\omega+i\zeta-\Delta_{\alpha\gamma}^{(m)}}
 +
 \frac{j_{\alpha\gamma}^{a(m)}j_{\gamma\alpha}^{b(-m)}}
      {\omega+i\zeta+\Delta_{\alpha\gamma}^{(m)}}
 \right],
 \label{eq:floquet-kubo-conductivity}
\end{align}
where $a,b\in\{x,y\}$, $f_{\alpha\mathbf{k}}$ is the occupation of Floquet band $\alpha$, $\zeta>0$ is a broadening parameter, and $N_{\mathbf{k}}$ is the number of momenta used to sample the Brillouin zone. We set $e=\hbar=1$ in the calculation and report the conductivity in units of $e^2/h$. The indices $\alpha$ and $\gamma$ label the inequivalent Floquet bands chosen within the first Floquet Brillouin zone, while $m$ accounts for transitions involving their replicas. The Fourier-resolved current matrix elements are
\begin{equation}
 j_{\alpha\gamma}^{a(m)}(\mathbf{k})
 =\frac{1}{T}\int_0^T\!dt\,
 e^{im\Omega t}
 \left\langle u_{\alpha\mathbf{k}}(t)\right|
 \frac{\partial H(\mathbf{k},t)}{\partial k_a}
 \left|u_{\gamma\mathbf{k}}(t)\right\rangle .
 \label{eq:floquet-current-matrix-element}
\end{equation}
Equivalently, after expanding $\lvert u_{\alpha\mathbf{k}}(t)\rangle =\sum\limits_l e^{-il\Omega t}\lvert u_{\alpha\mathbf{k}}^{(l)}\rangle$ and $j_a(\mathbf{k},t)=\sum_p e^{-ip\Omega t}j_a^{(p)}(\mathbf{k})$, Eq. \eqref{eq:floquet-current-matrix-element} becomes
\begin{equation}
 j_{\alpha\gamma}^{a(m)}(\mathbf{k})
 =\sum_{l,p}
 \left\langle u_{\alpha\mathbf{k}}^{(l)}\right|
 j_a^{(p)}(\mathbf{k})
 \left|u_{\gamma\mathbf{k}}^{(l+m-p)}\right\rangle .
 \label{eq:floquet-current-fourier}
\end{equation}
This expression can be evaluated directly using the truncated Floquet eigenvectors. For the results presented, we assume an ideal zero-temperature occupation in which the Floquet bands below the central quasienergy gap are filled and those above it are empty. Different nonequilibrium occupations may modify both the finite-frequency response and the dc limit.

Figures~\ref{OC-Fig-05} and \ref{OC-Fig-4d} show the real and imaginary parts of $\sigma^{xy}(\omega)$ for representative phases of Construction-I and Construction-II, respectively. The resonant structures arise from photon-assisted transitions satisfying \[ \omega\simeq \left|\Delta_{\alpha\gamma}^{(m)}(\mathbf{k})\right| \] at momenta where the corresponding current matrix elements are appreciable. They include direct interband transitions with $m=0$ and transitions involving other Floquet replicas with $m\neq0$. Their positions and spectral weights depend on both the quasienergy differences and the current matrix elements.

For completely filled or empty Floquet bands and in the clean dc limit $\zeta\rightarrow0^{+}$, the dc Hall conductivity approaches the limit that is determined by the Chern number of the occupied Floquet bands: 
\begin{equation} 
\lim_{\omega\rightarrow0} \operatorname{Re}\sigma^{xy}(\omega) 
=C_L\frac{e^2}{h}. \label{eq:dc-hall-chern} 
\end{equation} 
Under the ideal full-band occupation, the clean dc limits of the two $C_L=-2$ phases are therefore both $-2e^2/h$, although their winding numbers are $(W_0, W_{\pi})=(0,+2)$ and $(-2,0)$. The dc Hall response consequently determines $C_L=W_0-W_{\pi}$ but does not separately identify the two gap winding numbers. Their gap-resolved topology must instead be determined from the winding invariants and edge spectra.

\section{Possible experimental realizations}
\label{Sec:Experiment}

%\subsection{Photonic waveguide arrays}
Possible implementations of the static $N$-stacked SSH lattice in optical-waveguide arrays, electrical circuits, and magnetomechanical systems were discussed in Ref. \cite{Agrawal_2022-02}.  To realize the complete model, additional requirements are the two-dimensional interchain couplings and periodic modulation discussed below. 

Multimode femtosecond-laser-written waveguides provide one possible photonic route to the driven $N$-stacked SSH model. The model requires co-located $A$ and $B$ orbitals and a circular drive. Two transverse modes centered on each waveguide could span the local $A-B$ orbital space, while couplings between different waveguides would generate the intersite hopping terms. In the paraxial approximation, the propagation coordinate $z$ plays the role of time. If the waveguides follow helical trajectories, transformation to the co-moving frame produces an effective circular vector potential \cite{Rechtsman2013},
\begin{equation}
\begin{aligned}
\bs{\mathcal{A}}(z)
&=k_0R\Omega_z
\left(\sin\Omega_z z,-\cos\Omega_z z\right),\\
k_0&=\frac{2\pi n_0}{\lambda},
\qquad
\Omega_z=\frac{2\pi}{Z}.
\end{aligned}
\label{Eq-photonic-drive}
\end{equation}
Up to a shift of the origin of $z$, the helix handedness written in  Eq. (36) corresponds to $\chi=+1$ in the convention of  Eq. (8). The choice $\chi=-1$ used in our
calculations is obtained by reversing the helix handedness. Here, $R$ and $Z$ are the helix radius and pitch, respectively, $n_0$ is the refractive index, and $\lambda$ is the optical wavelength. A coupling across the displacement $\bs{r}_{mn}$ then acquires the Peierls factor $\exp[i\bs{\mathcal{A}}(z)\boldsymbol{\cdot}\bs{r}_{mn}]$. If $a$ denotes the physical length corresponding to a unit bond of the model, the dimensionless Peierls amplitude is $A_0=k_0R\Omega_z a$. Using the parameters of Ref. \cite{Rechtsman2013}, $\lambda=633\,\mathrm{nm}$, $n_0=1.45$, $R=8\,\mu\mathrm{m}$, $Z=1\,\mathrm{cm}$, and $a=15\,\mu\mathrm{m}$, we obtain $A_0\simeq1.1$. This shows that Peierls amplitudes of the order considered in our phase diagrams are experimentally accessible.

In this orbital implementation, the local term $t_0(1-\eta)$ can be represented by the propagation-constant splitting between two co-centered orbital modes after a local basis rotation. The intersite hopping amplitudes can be controlled through the waveguide separations and orientations. Realizing the complete Hamiltonian requires engineering the magnitudes and relative signs of the orbital-changing vertical and diagonal couplings while suppressing same-orbital couplings in the rotated A–B basis. 

A related two-dimensional SSH weak topological phase has also been realized in an electrical-circuit platform \cite{Yang-2022}.
The driven model considered here additionally requires the co-located two-orbital embedding specified above and a circular drive. A direct static experimental precedent for such an embedding is the monoatomic mechanical lattice of Ref. \cite{Liu2024}, where each resonator hosts even- and odd-parity orbital modes at the same effective lattice site. Interorbital coupling between neighboring waveguides has also been demonstrated \cite{GuzmanSilva2021}. These experiments establish the required orbital ingredients. 
Possible cold-atom and electronic implementations of the driven model, together with their requirements and limitations, are discussed in Appendix \ref{App:B}

\section{\label{Sec:Discussion}Summary}

In summary, we investigate the Floquet topology of a two-dimensional lattice constructed by stacking one-dimensional SSH chains and subjecting it to circularly polarized light. We compare two constructions formed from constituent SSH chains that are topologically trivial (Construction-I) or nontrivial (Construction-II). Using a truncated Floquet-Sambe representation, we calculate the quasienergy spectrum and characterize the driven phases through the Floquet-band Chern number $C_L$ and the directional Zak phases $(Z_x, Z_y)$. The phase diagrams demonstrate that the two constructions develop distinct topological structures under the same driving protocol. Construction-I supports phases with $C_L=0,\pm1,-2,$ and $-3$, whereas Construction-II supports $C_L=0,\pm1,-2,$ and $-4$. In both cases, phases with $|C_L|>1$ occur in the low-frequency regime, while the high-frequency regime is dominated by phases with $|C_L|=1$.

The Chern number alone, however, does not determine how the topology is distributed between the central and Floquet-zone-edge quasienergy gaps. We therefore classify representative phases using the dynamical winding numbers $(W_0,W_{\pi})$, which satisfy $C_L=W_0-W_{\pi}$. For Construction-I, the phases characterize by $(W_0,W_{\pi})=(0,+2)$ and $(-1,+2)$, giving $C_L=-2$ and $-3$, respectively. For Construction-II, the corresponding representative phases have $(W_0,W_{\pi})=(-2,+2)$ and $(-2,0)$, giving $C_L=-4$ and $-2$. The two phases, therefore, provide the most direct manifestation of construction dependence with the same Chern number $C_L=-2$: the Construction-I phase carries its net chirality entirely in the Floquet-zone-edge gap, whereas the Construction-II phase carries it entirely in the zero-quasienergy gap. The cylindrical spectra for both boundary orientations are consistent with this gap-resolved Floquet bulk-boundary correspondence.

As a representative bulk consequence of these driven phases, we also examine the time-averaged optical Hall response for both constructions. For the ideal occupation used here, the dc Hall conductivity is given by $C_L e^2/h$, whereas the finite-frequency response includes photon-assisted transitions between different Floquet sectors. Consequently, the Hall response probes the bulk Floquet Chern character but does not determine $W_0$ and $W_\pi$ separately. The winding invariants and the edge spectra are therefore essential for identifying whether the chiral topology resides in the zero or $\pi$ quasienergy gap. The two constructions, therefore, provide distinct realizations of Floquet topology despite being equivalent in the static limit. Their difference originates from the bond-geometry-dependent Peierls dressing of the hopping processes under circular driving. In particular, the same Floquet-band Chern number can be distributed differently between the zero and Floquet-zone-edge gaps.

\begin{acknowledgments}
A.A.  thanks the Center for Advanced Computation (CAC) at Korea Institute for Advanced Study (KIAS) for providing computing resources for this work. This work is supported by the KIAS Individual Grants, PG098801 (A.A.). JNB acknowledges the Department of Science and Technology (DST) for providing computational resources through the FIST program (Project No. SR/FST/PS-1/2017/30).
\end{acknowledgments}

\appendix
\section{Symmetry classification of the undriven stacked-SSH model}
\label{app:static_symmetry}

Since the Bloch Hamiltonian contains only $\sigma_x$ and $\sigma_y$ components, it obeys spinless time-reversal, particle-hole, and chiral symmetries. The corresponding operators are
\begin{equation}
\mathcal{T}=\mathcal{K},\qquad
\mathcal{C}=\sigma_z\mathcal{K},\qquad
\Gamma=\sigma_z,
\end{equation}
where $\mathcal{K}$ denotes complex conjugation. They satisfy
\begin{align}
\mathcal{T}H(\mathbf{k})\mathcal{T}^{-1}
 &=H(-\mathbf{k}),\\
\mathcal{C}H(\mathbf{k})\mathcal{C}^{-1}
 &=-H(-\mathbf{k}),\\
\Gamma H(\mathbf{k})\Gamma^{-1}
 &=-H(\mathbf{k}).
\end{align}
Moreover, $\mathcal{T}^2=\mathcal{C}^2=+1$. The undriven Hamiltonian therefore belongs to symmetry class BDI.

\section{Other Experimental ideas}
\label{App:B}
\subsection{Cold atoms in shaken optical lattices}

Dimerized tunneling along the chain direction has been realized in optical superlattices \cite{Atala2013}. However, these realizations generally employ two spatially separated sites within each unit cell and therefore do not directly implement the co-located orbital embedding assumed here. A cold-atom realization of the present model would instead require two internal states with a common spatial wave function at each Bravais-lattice site, coupled on site by a spatially uniform (e.g., radio-frequency) field,
together with engineered state-changing intersite tunneling, including the vertical and diagonal couplings and their relative signs.

Circular lattice shaking can generate the periodic momentum shift used in the driven Hamiltonian \cite{Jotzu2014}. Denoting the laboratory angular frequency by $\omega_{\rm sh}$, the dimensionless parameters used in our calculations are
\begin{equation}
\Omega=\frac{\hbar\omega_{\rm sh}}{t_0},
\qquad
A_0=\frac{M\omega_{\rm sh}d a}{\hbar},
\label{Eq-shaking-parameters}
\end{equation}
where $M$ is the atomic mass, $d$ is the shaking radius, and $a$ is the Bravais-lattice constant.

As a reference experimental scale, we use the $^{40}$K optical superlattice of Ref. \cite{Viebahn2024}, which has a two-site unit-cell length of $1.064~\mu{\rm m}$ and an average tunneling of $t/h=368(25)$~Hz. Identifying these as representative values for $a$ and $t_0$, respectively, the point $(A_0,\Omega)=(0.5,2.5)$ considered in our calculations corresponds to
\begin{equation}
\frac{\omega_{\rm sh}}{2\pi}\simeq0.92~{\rm kHz},
\qquad
d\simeq0.13~\mu{\rm m}.
\label{Eq-cold-atom-estimate}
\end{equation}
These values provide an illustrative conversion to laboratory scales; Ref. \cite{Viebahn2024} does not realize the complete two-dimensional driven model considered here.

The Berry-curvature response of the Floquet bands could be probed through the transverse center-of-mass motion induced by a weak constant force, following the approach of Ref. \cite{Jotzu2014}. Such atomic transport is analogous to a Hall response, but it is not a direct measurement of the frequency-dependent electronic optical conductivity calculated in Sec. \ref{Sec:OC}.

\subsection{Electronic systems}

For an electronic implementation, let $\omega_{\rm d}$ denote the physical angular frequency of the applied field. The dimensionless parameters satisfy
\begin{equation}
\Omega=\frac{\hbar\omega_{\rm d}}{t_0},
\qquad
A_0=\frac{eaE_0}{\hbar\omega_{\rm d}},
\label{Eq-electronic-parameters}
\end{equation}
where $E_0$ is the peak electric-field amplitude and $e$ is the magnitude of the electronic charge. Therefore,
\begin{equation}
E_0=\frac{A_0\Omega t_0}{ea}.
\label{Eq-electronic-field}
\end{equation}
The required field consequently scales as $t_0/a$, favoring narrow-band systems with relatively large unit cells.  
A quantitative estimate would require identifying a material with two co-located internal orbitals at each Bravais-lattice site that also realizes the specific intrachain, vertical, and diagonal hopping pattern of the model. Heating, dissipation, and nonequilibrium Floquet-band occupations would also modify the ideal-occupation Hall response calculated in Sec. \ref{Sec:OC}. Equation \eqref{Eq-electronic-field} should therefore be regarded as a general scaling relation rather than a material-specific feasibility estimate.

\bibliography{References}

\end{document}